\documentclass{amsart}
\usepackage{graphicx}
\newtheorem{lem}{Lemma}

\newtheorem{thm}{Theorem}
\newtheorem{prop}{Proposition}

\newtheorem{ex}{Example}
\newtheorem{cor}{Corollary}
\newcommand{\pf}{{\noindent\bf Proof:\,}}

\newcommand{\w}{\omega}
\newcommand{\F}{\mathbb{F}}
\newcommand{\Z}{\mathbb{Z}}

\begin{document}

\title{Cyclic codes over $M_2(\F_2)$}


\author{Adel Alamadhi$^*$}
\address{$*$MECAA, King Abdulaziz University, Jeddah, Saudi Arabia.}
\email{adelnife2@yahoo.com}

\author{Houda Sboui}
\address{ ENIT, El Manar, Tunis, Tunisia}
\email{sboui.houda@yahoo.fr}
\author{Patrick Sol\'e$\ddag,^*$}
\address{$\ddag$Telecom ParisTech, 46 rue Barrault, 75634 Paris Cedex 13, France.}
\email{sole@telecom-paristech.fr}
\author{Olfa Yemen}
\address{ Institut Preparatoire, El Manar, Tunis, Tunisia}
\email{olfa{\_}yemen@yahoo.fr}

\begin{abstract}
The ring in the title is the first non commutative ring to have been used as alphabet for block codes. The original  motivation was the construction of some quaternionic modular lattices from codes. The new application is the construction of  space time codes obtained by concatenation from the Golden code. In this article, we derive structure theorems for cyclic codes over that ring, and use them to characterize the lengths where self dual cyclic codes exist. These codes  in turn give rise to formally self dual quaternary codes.
\end{abstract}

\maketitle
\section{Introduction}
Since its inception in the late forties of the last century, algebraic coding theory was confined to using finite fields as alphabets. Then, in the early nineties, partly motivated by engineering applications, like spreading sequence design for CDMA, came a surge of interest for codes over rings
that crystallized in the prized paper \cite{HK+}. That work received the Best Paper award for 1994 from the IEEE Information Theory Society, because it solved a twenty year old riddle in coding theory, the formal duality of Kerdock and Preparata codes. These two infinite families of {\em nonlinear} binary codes are connected by a MacWiliams relation at the weight enumerator level. The key to unlocking this mystery was to use duality of codes over $\Z_4,$ a ring of order $4$ that is not a finite field. Since then the number of publications in the field of codes over ring has exploded and the cited  paper has been quoted since more than $720$ times as per Google scholar.

Still, from 1994 till now there have been very few papers on codes over {\bf non commutative rings}. The first concrete such alphabet seems to have been $A=M_2(\F_2)$, which appeared in algebraic constructions of modular lattices \cite{B}. This alphabet resurfaced recently in connection with the new topic of space time codes \cite{OSB}.
The beautiful fact about this alphabet is a Gray map analogue of that of \cite{HK+}, the Bachoc map  that maps isometrically this ring of order $16$ with a special distance we call the Bachoc distance onto two copies of the Galois Field $\F_4,$ with the Hamming distance \cite[\S 6.2]{B}. The idea is to introduce two matrices allusively called $\omega$ and $i$ such that
their respective characteristic polynomials are $X^2+X+1$ and $X^2+1$ and satisfying the relation $i\omega =\omega^2 i.$ The ring $A$ can be written as $A=\F_4+i\F_4,$ by regarding $\F_4$ as $\F_2 [\omega].$ This confers to it a quotient structure over a skew polynomial ring with coefficient ring the field $\F_4.$ Note, for completeness, that if $M$ is a nonzero matrix of $A$ then its Bachoc weight is worth $2$ if $M$ is singular nonzero, $1$ if $M$ is regular \cite{B}.

 It seems legitimate, following a long trend in research to apply the methodology of cyclic codes over that simple and maybe simplest example of non commutative finite ring. In this article we characterize cyclic
codes by their generators. Their duals are also cyclic and their generators can be expressed simply as a function of the generators of the primal codes. We give an arithmetic criterion for a cyclic code to be self dual for the Euclidean scalar product. We show that self dual cyclic codes for the Hermitian scalar product cannot exist in odd length. We show that the Bachoc image of a self dual code for the Euclidean scalar product is formally self dual. Our new expression of the Bachoc image as a Plotkin sum of the residue and torsion codes allows us to compute the parameters of many examples of formally self dual quaternary codes for $n\le 31.$

\section{Notation and definitions}
For simplicity, let $A=M_2(\F_2)$ the ring of matrices of order $2$ over the finite field $\F_2.$ Following \cite{B} write
$$ A=\F_2[\omega]\oplus i \F_2[\omega],$$ where $\omega$ and $i$ are in $A,$ and are such that
their respective characteristic polynomials are $X^2+X+1$ and $X^2+1$ and satisfying the relation $i\omega =\omega^2 i.$ A possible choice is
\begin{displaymath}
i=
 \setcounter{MaxMatrixCols}{2}
 \begin{pmatrix}
0& 1\\
1&0\\

\end{pmatrix}  ,\; \omega=
     \setcounter{MaxMatrixCols}{2}
 \begin{pmatrix}
0& 1\\
1&1\\

\end{pmatrix}.
\end{displaymath}
For convenience we let $u=1+i,$ a nilpotent element, identify the subring $ \F_2[\omega]$
with $\F_4$ and write
$$ A=\F_4\oplus u \F_4.$$

Now, denote by $\mu$ the projection on the first component of that direct sum. Note that this map is $\F_4- $ linear but not a ring morphism.This map extends coefficient wise to a map from $A[X]$ down to $\F_4[X].$ Given an $A-$ linear code $C$
of length $n,$ that is a $A$ submodule of $A^n$ we construct two linear quaternary codes of the same length $,n$  denoted by $R$ and $T$ , for {\bf  residue} and {\bf torsion code} respectively, such that $R=\mu(C)$ and $T$ is the largest quaternary code $D$ with the property that $uD \subseteq C.$ We see by considering $uC,$ that $R\subseteq T,$ with equality iff $C$ is a free $A-$module.

Let $R_n=A[X]/(X^n -1)$ denote the ring whose right sided  ideals represent cyclic codes of length $n.$
\section{Structure theorems}
We prepare for the proof of Theorem 1  by a pair of Lemmas. Consider the product
$x^n-1=\prod_{j=1}^t f_j,$ where the $f_j$'s are irreducible polynomials over $\F_4.$
{\em In the whole paper we assume $n$ to be odd}. Hence these polynomials are pairwise distinct.
We need a non commutative analogue of the CRT for modules rather than ideals.

{\lem As right modules we have the expansion

$$ R_n=\oplus_{j=1}^t A_j,$$
where the $A_j=A[X]/(f_j)$ are quotient $A-$modules.}

\pf The proof is an immediate application of \cite[9.12]{W}.
\qed

Note that if $(f_j)$ is not two sided then $A_j$ is not a ring, but only a right $A-$module.
Further, we have an analogue of  Lemma $3$ of \cite{PQ} where $\F_4$ plays the role of $\F_2$ wrt to $A$ which plays the role of $\Z_4.$ Note that since $\F_4$ is a subring of $A,$ Hensel
lifting is trivial in our context. For self containment we rederive the proof.

{\lem If $f$ is an irreducible polynomial over $\F_4$ the only right $A-$modules of $R(f)=A[X]/(f)$ are
$(0),\,(u),\,(1).$ In particular this quotient ring is a non commutative chain ring. }

\pf
Let $I\neq (0)$ be an ideal of $R(f)$. Pick a $g$ in $A[X]$ such that $g+(f) \in I,$ but $g \notin (f).$ Because $f$ is irreducible the GCD of $\mu g$ and $f$ can only take two values $1$ or $f.$
In the first case $g$ is invertible mod $f$ and $I=(1)=R(f).$ If this never happens $I \subseteq u+(f).$ Let us show the reverse inclusion. Let $g=ur$ with $ur+(f) \subseteq I$ and $ur+(f)\neq 0.$
We can assume by the latter condition that $\mu r \notin (f).$ Hence by irreducibility of $f$ we see that $GCD(\mu r,f)=1.$ This entails the existence of $a,b,c \in A[X]$ such that
$$ ra +fb=1+uc, $$ and, multiplying both sides  by $u$ that $$ura=u+ufb.$$
The LHS is in $I,$ a right sided ideal. The reverse inclusion follows.
\qed

We are now ready for the main result of this section.

{\thm Given a factorization $X^n-1=fgh$ into three pairwise coprime factors over $\F_4[X]$  we can construct the cyclic code
$$ C=(fh)+u(fg).$$  The residue and torsion codes of $C$  are cyclic quaternary codes of length $n$ with
respective generator polynomials $fh$ and $f,$ of respective dimensions $deg(g)$ and $deg(g)+deg(h)$ over $\F_4.$ Conversely any cyclic $A-$code of length $n$  arises in this way.}

\pf The code so constructed is a right ideal of $R_n,$ because $(fg)\subseteq (f).$ In the other
direction we combine the two above lemmas to see that every cyclic code is a sum of ideals of
some $A_j$'s some of the form $(\widehat{f_j})$ some of the form $u( \widehat{f_j})$ where
the $f_j$ are as in Lemma 1 and where we let $\widehat{f_j}=(X^n-1)/f_j.$

\qed

Thus any cyclic code is obtained by a "multilevel construction." This situation is different from that over $\Z_4$
but similar to what happens over $\F_2+u\F_2,$ where like for $A$ the residue field is a subring \cite{BU}.
\section{The Bachoc map}
\subsection{Metric properties}
Define the  Bachoc map of $a+bi$ with $a,\,b \in \F_4$ by the formula
$$\phi(a+ib)=(a,b).$$
Alternatively by using $u=1+i$ we see that
$$\phi(a+ub)=(a+b,b).$$
By extending this map componentwise to vectors of $A^n$ we see a connection with the
$(u,u+v)$ construction, also known as Plotkin sum of two codes \cite{P}. If $C_1,\,C_2$ are two
quaternary codes of length $n$ the Plotkin sum is defined as
$$C_1 P C_2=\{(u,u+v) \vert \,u\in C_1,\, v\in C_2 \}.$$
Given their  dimensions $k_1,\,k_2$ and their  distances $d_1,\,d_2,$ it is well-known that
the parameters of the Plotkin sum are $[2n,k_1+k_2,\min(2d_1,d_2)].$
The following proposition seems to have been unnoticed so far.
{\prop The Bachoc map of a cyclic code is equivalent to  the Plotkin sum of its   torsion and residue code.  }

\pf By Theorem 1 we know that for such a cyclic code

$$C=R+uT$$
where $R$ and $T$ are its residue and torsion code. The result follows by definition of the
Bachoc map.
\qed
\subsection{Duality properties}
Following \cite[Prop. 2.1 (2)]{B} we define a {\bf conjugation} on $A$ by the rule
$$\overline{a+ib}=\overline{a}+i b,$$
valid for $a,\, b \in \F_4,$ and where the bars in the RHS are for conjugation in $\F_4,$
ie $\overline{a}=a^2.$
Extending this to vectors we can define an hermitian form $\sum_j x_j \overline{y_j}$  on $A^n$ and  obtain by \cite[Lemma 6.4]{B} the following useful result, which means that the Bachoc map is compatible with hermitian duality on range and domain.

{\prop If $C \subseteq A^n$ is self orthogonal for the above hermitian form then $\phi (C)$ is self orthogonal for the classical quaternary hermitian form on $\F_4^{2n}.$}

\pf Follows immediately from the identity
$$(a+bi)\overline{(a'+b'i)}=a  \overline{a'}+b \overline{b'}+(ba'+ab')i. $$

 \qed

We now consider self duality wrt the euclidean form $\sum_j x_j{y_j}$  on $A^n.$ Define the
{\bf Bachoc weight enumerator} of $C\subseteq A^n$ by
$$bwe_C(a,b,c)=\sum_{c \in C} a^{n_0(c)}b^{n_1(c)}c^{n_2(c)}$$
where $n_j(c)$ is the number of entries in $c$ of Bachoc weight $j.$ Recall that for a quaternary code $Q$  of length $N$ the {\bf Hamming weight enumerator} is the bivariate homogeneous polynomial
$$W_Q(x,y) =\sum_{q \in Q} x^{N-|q|}y^{|q|}.$$ In particular a quaternary code is self dual for the hamming weight enumerator iff that polynomial is a fixed point of the MacWilliams transform or in symbols
$$W_Q(x,y)=W_{Q}(\frac{x+3y}{2},\frac{x-y}{2}).$$
 The next result is an analogue of a result of \cite{HK+}, which is the key to the formal duality of Kerdock and Preparata codes.
{\prop If $C \subseteq A^n$ is self dual for the euclidean  form then $\phi (C)$ is
formally self dual for the Hamming weight enumerator.}

\pf The Hamming weight enumerator of  $\phi (C)$ is obtained from the Bachoc weight enumerator of $C$ as
$$W_{\phi (C)}(x,y)=bwe_C(x^2,xy,y^2).$$
By MacWilliams relation ( \cite[Th. 4.2 (2)]{B}) on $A^n$  we can
express the bwe of $C$ as a function of the bwe of the dual by
$$ bwe_C(a,b,c)=\frac{1}{|C|}  bwe_C(a+6b+9c,a+2b-3c,a-2b+c) $$
and eliminating the bwe's by $a=x^2,\,b=xy,\,c=y^2$
we get, using the homogeneity 	and $|C|=4^n=2^{2n},$ the identity
$$W_{\phi (C)}(x,y)=W_{\phi (C)}(\frac{x+3y}{2},\frac{x-y}{2})$$
as we should.

 \qed
 \subsection{Cyclicity properties}
 The cyclicity of the Bachoc image of a cyclic code will result from a structure theorem on repeated root cyclic codes
 over $\F_4.$
 {\lem\label{vl} Assume $C_1$ and $C_2$ are quaternary codes of odd length $n$ and generators polynomials $g_1$ and $g_1g_2$. The cyclic code of length $2n$ and generator $g_1^2g_2$ is equivalent to the Plotkin sum of $C_1$ and $C_2.$ }
 The proof is a straightforward extension of the proof of Theorem 1 in \cite{v} and is omitted.
 {\prop The Bachoc image of a cyclic code over $A$ of odd length $n$ is equivalent to a cyclic code of length $2n$
 of generator $g_T^2g_R$ where $g_T$ and $g_R$ are the generators of its residue and torsion code, respectively.}
 \pf
 Combine Lemma \ref{vl} with Proposition 1.

 \qed
\section{Self dual cyclic codes}
First, we characterize the (euclidean) dual of a cyclic code by its generators. Denote by $ f^*$
the reciprocal polynomial of $f,$ made monic after normalization. For instance $(x+\omega)^*=x+\omega^2.$
{\lem Let $C=(fh,ufg)$ be a cyclic code of odd  length $n$ with $X^n-1=fgh.$ The dual of $C$
is $C^{\perp}=(g^*h^*, ug^* f^*).$  }

\pf  Inclusion of the RHS in the LHS is easy to check. Equality follows by dimension count.
\qed

{\thm \label{sd}  Let $C=(fh,ufg)$ be a cyclic code of odd  length $n$ with $X^n-1=fgh.$  This code is Euclidean self dual iff $h= h^*$ and $g= f^*.$ It is never Hermitian self dual.}

\pf By the preceding Lemma the condition for Euclidean self duality  is sufficient. To see necessity identify generators
$$fh=g^*h^*$$
and
$$fg=g^*f^*. $$
Multiplying the first equality by $g$ and the second by $ h$ we get
$$gh^*=hf^*.$$
Since $h$ is coprime with $g$ and of the same degree as $h^*$ we see that $h=\epsilon h^*$ for some $\epsilon=1,\omega,\omega^2 .$
But because $x+1$ must divide $h$ we see that $\epsilon=1.$
The same line of reasoning in the Hermitian case would lead to $f=g$ which is impossible for $n$ odd.
\qed

The existence of a triple of polynomials satisfying the above theorem yields some non trivial consequences on cyclotomic cosets.
Recall that the multiplicative order of some integer $a$ modulo $b$ is the smallest $j$ such that $a^j =1 \pmod b.$
{\cor \label{ari}Non trivial self dual  cyclic codes of length $n$  exist iff there is no $j$ such that $4^j =-1 \pmod n,$ or, equivalently if the multiplicative order of $4$ mod $n$ is odd.}

\pf
If $-1$ is a power of $4$ modulo $n$  then all 4-cyclotomic cosets are symmetric and there are no nontrivial
$f$ and $g$ satisfying the Theorem hypothesis. Conversely, if there are non symmetric cyclotomic cosets, let $\Z_n$ be split into $T \bigcup -T \bigcup U,$ with $U=-U$ with $T$ the union of these.
Take $f$ to be the polynomial whose roots correspond to $T$ and $h$ that whose roots correspond to $U.$ The cyclic code attached to the polynomial triple $f, \, f^*,\, h$ is self dual, and non trivial
since $f$ is.
\qed

{\ex If $n=5$ we see that $4 \equiv -1 \pmod{5},$ and there are no non trivial cyclic codes. Indeed the factorization of $$X^5+1=(X+1)(X^2+\omega x+1)(X^2+\omega^2X+1)$$ is into three self reciprocal polynomials.}

The appendix of \cite{PSQ} contains a detailed study of the weaker condition $n$ divides $2^k+1$ for some $k.$ Let $N(x)$ denote the number of primes $p\le x$ such that the multiplicative order of $4$ mod $p$ is odd. By \cite[Th. 1]{Mo2} we know that for $x$ large we have
$$N(x) \sim \frac{7x}{12 \log x} .$$
In particular there exists arbitrarily long non trivial self dual cyclic codes over $A.$
\section{Self dual cyclic codes of odd  length $n\le 31.$}
In the following we classify self dual cyclic codes taking into account the symmetry between $f$ and $g$ in Theorem \ref{sd}. Since $g=f^*,$ swappping $f$ and $g$ leads to equivalent codes up to coordinate order reversion.
Note that $h$ has to be divisible by the product of all the irreducible self reciprocal polynomials that divide $x^n+1.$
\begin{itemize}
\item[$n=3$] Two cyclic self dual codes with $h=X+1$ and $f,\, g=X+\omega.$

\begin{table}[ht]
\small
\begin{tabular}{|l|l|}
\hline
$h$ & $x+1$  \\ \hline
$f$ & $x+w$  \\ \hline
$d_R$ &3 \\ \hline
$d_T$ &2 \\ \hline
$\min(2d_T,d_R)$&3 \\ \hline
\end{tabular}
\label{table8}

\caption{ Length $3$ }
\end{table}

\item[$n=5$] As seen before there are no nontrivial  cyclic self dual  codes by Corollary \ref{ari}.
\item[$n=7$] Two cyclic self dual codes with $h=X+1$ and $f,\, g=X^3+X+1.$

\begin{table}[ht]
\small
\begin{tabular}{|l|l|}
\hline
$h$ & $x+1$  \\ \hline
$f$ & $x^3+x+1$  \\ \hline
$d_R$ &4 \\ \hline
$d_T$ &3 \\ \hline
$\min(2d_T,d_R)$&4 \\ \hline
\end{tabular}
\label{table8}

\caption{ Length $7$ }
\end{table}

\item[$n=11$] Two cyclic self dual codes with $h=X+1$ and $f,\, g=X^5+X^5 + \w X^4 + X^3 + X^2 + \w^2 X + 1  .$

\begin{table}[ht]
\small
\begin{tabular}{|l|l|}
\hline
$h$ & $x+1$  \\ \hline
$f$ & $x^5+w*x^4+x^3+x^2+w^2*x+1$  \\ \hline
$d_R$ &6 \\ \hline
$d_T$ &5 \\ \hline
$\min(2d_T,d_R)$&6 \\ \hline
\end{tabular}
\label{table8}

\caption{ Length $11$ }
\end{table}

\item[$n=13$]No cyclic non trivial self dual cyclic codes by Corollary \ref{ari}, since $13$ divides $4^3+1.$
\item[$n=15$] The factorization of $X^n+1$ is of the form $(x^5+1)f_1f_1^*f_2f_2^*f_3f_3^*,$ with $f_1=x^2+x+\omega,$ $f_2=x^2+x+\omega^2,$ and
$f_3=x+\omega.$ We discuss according to the number of factors  of $h.$
\begin{itemize}
\item  If we take $h=x^5+1$ then we have four choices for $f,\,g$ namely $f=f_1f_2f_3,\,f=f_1f_2f_3^*,\,f=f_1f_2^*f_3,\,f=f_1f_2^*f_3^*.$
\item If we take $h=(x^5+1)f_1f_1^*$ then we have two choices for $f,\,g$ namely $f=f_2f_3,\,f=f_2f_3^*$ and similarly two choices for $f,\,g$ in the case of
 $h=(x^5+1)f_2f_2^*$, and again two choices for $f,\,g$ in the case of  $h=(x^5+1)f_3f_3^*$
\item  If we take $h=(x^5+1)f_1f_1^* f_2f_2^*$ then we have only one choice for $f$ that is $f=f_3,$ and similarly for  $h=(x^5+1)f_1f_1^* f_3f_3^*$ and  $h=(x^5+1)f_2f_2^* f_3f_3^*$
\end{itemize}

\begin{table}[ht]
\small
\begin{tabular}{|l|l|l|l|l|l|l|l|l|l|l|}
\hline
$h/(x^5+1)$ &$1$ & $1$ & $1$ & $1$& $f_1f_1^*$ &$f_1f_1^*$ &$f_2f_2^*$  &$f_2f_2^*$  &$f_3f_3^*$& $f_3f_3^*$ \\ \hline
$f$ &$ f_1f_2f_3$ & $f_1f_2f_3^*$&$ f_1f_2^*f_3$ & $ f_1f_2^*f_3^*$& $f_2f_3$  & $f_2f_3^*$&$f_1f_3$  & $f_1f_3^*$&$f_1f_2$  & $f_1f_2^*$\\ \hline
$d_R$ &8&8&6&3 &9&11&11 &9&8 & 6\\ \hline
$d_T$&3&3&3&2  &2&3&3 &2&3&2 \\ \hline
$\min(2d_T,d_R)$&6&6&6&3  &4&6&6 &4&6&4 \\ \hline
\end{tabular}
\label{table1}

\caption{ Length $15$}
\end{table}
\begin{table}[ht]
\small
\begin{tabular}{|l|l|l|l|}
\hline
$h/(x^5+1)$ & $f_1f_1^*f_2f_2^*$   & $f_1f_1^*f_3f_3^*$ &  $f_2f_2^*f_3f_3^*$\\ \hline
$f$ & $f_3$ & $f_2$& $f_1$ \\ \hline
$d_R$ &15&12&12 \\ \hline
$d_T$ &2&2& 2\\ \hline
$\min(2d_T,d_R)$&4&4&4 \\ \hline
\end{tabular}
\label{table2}

\caption{ Length $15$ continued}
\end{table}
\item[$n=17$] No non trivial self dual cyclic codes by Corollary \ref{ari},  since $17$ divides $4^2+1.$
\item[$n=19$] Two cyclic self dual codes with $h=X+1$ and $f,\, g=X^9+ \w X^8 + \w X^6 + \w X^5 +\w^2X^4 +\w^2X^3 +\w^2X + 1.$
\item[$n=21$] The factorization of $X^n+1$ is of the form  $(X^3+1)f_0f_0^*f_1f_1^*f_2f_2^*,$ with $f_i=X^3+\w^iX+1.$ If $h$ is a multiple of $X^3+1,$ the discussion is the same as for length $15.$ If $h=X+1$ we take $f=(X+\w)f_0^af_1^bf_2^c$ with $a,b,c$ in $\{1,*\}.$

\begin{table}[ht]
\small
\begin{tabular}{|l|l|l|l|l|l|l|l|l|l|l|}
\hline
$h/(x^3+1)$ &$1$ & $1$ & $1$ &$1$ &$f_0f_0^*$ &$f_0f_0^*$ &$f_1f_1^*$  &$f_1f_1^*$  &$f_2f_2^*$& $f_2f_2^*$\\ \hline
$f$ &$ f_0f_1f_2$ & $f_0f_1f_2^*$ &$ f_0f_1^*f_2$& $ f_0f_1^*f_2^*$& $f_1f_2$  & $f_1f_2^*$&$f_0f_2$  & $f_0f_2^*$&$f_0f_1$  & $f_0f_1^*$\\ \hline
$d_R$ &4 &6&6&8&8&6&8&12&8 &12  \\ \hline
$d_T$&3&3&3&5 &3&2&3&3&3 &3\\ \hline
$\min(2d_T,d_R)$&4&6&6&8 &6&4&6&6&6 &6\\ \hline
\end{tabular}
\label{table3}

\caption{ Length $21$}

\end{table}

\begin{table}[ht]
\small
\begin{tabular}{|l|l|l|l|}
\hline
$h/(x^3+1)$ & $f_0f_0^*f_1f_1^*$   & $f_0f_0^*f_2f_2^*$ &  $f_1f_1^*f_2f_2^*$\\ \hline
$f$ & $f_2$ & $f_1$& $f_0$ \\ \hline
$d_R$ &12&12&12 \\ \hline
$d_T$ &2&2&2 \\ \hline
$\min(2d_T,d_R)$&4&4&4 \\ \hline
\end{tabular}
\label{table4}

\caption{ Length $21$ continued}
\end{table}
\begin{table}[ht]
\small
\begin{tabular}{|l|l|l|l|l|l|l|l|l|}
\hline

$abc$ & $111$ & $11*$& $1*1$ & $1**$& $*11$& $*1*$& $**1$& $***$\\ \hline
$d_R$ &4&6&3&8&8&6&3&4 \\ \hline
$d_T$ &4&6&3&5&5&6&3&4 \\ \hline
$\min(2d_T,d_R)$  &4&6&3&8&8&6&3&4 \\ \hline
\end{tabular}
\label{table4}

\caption{ Length $21$ and $h=X+1$}
\end{table}
\item[$n=23$] Two cyclic self dual codes with $h=X+1$ and $f,\, g=X^{11}+X^9+X^7+X^6+X^5+X+1  .$
\item[$n=25$]No cyclic non trivial self dual cyclic codes  by Corollary \ref{ari},  since $25$ divides $4^5+1.$
\item[$n=27$]  The factorization of $X^n+1$ is of the form $ (x+1)f_1f_1^*f_2f_2^*f_3f_3^*,$ with $f_1=x^1+\w,\,f_2=x^3+\w,\,f_3=x^9+\w.$ Again the discussion is the same as for $n=15.$
\begin{table}[ht]
\small
\begin{tabular}{|l|l|l|l|l|l|l|l|l|l|l|}
\hline
$h/(x+1)$ &$1$ & $1$ & $1$& $1$ & $f_1f_1^*$ &$f_1f_1^*$ &$f_2f_2^*$  &$f_2f_2^*$  &$f_3f_3^*$& $f_3f_3^*$ \\ \hline
$f$ &$ f_1f_2f_3$ & $f_1f_2f_3^*$& $ f_1f_2^*f_3$ & $ f_1f_2^*f_3^*$& $f_2f_3$  & $f_2f_3^*$&$f_1f_3$  & $f_1f_3^*$&$f_1f_2$  & $f_1f_2^*$\\ \hline
$d_R$ &3&3&3& 3&3&3&3 &3&9&9  \\ \hline
$d_T$&3&3&3& 3 &3&3&3 &3&2&2 \\ \hline
$\min(2d_T,d_R)$&3&3&3& 3 &3&3&3 &3&4&4 \\ \hline
\end{tabular}
\label{table5}

\caption{ Length $27$}
\end{table}
\begin{table}[ht]
\small
\begin{tabular}{|l|l|l|l|}
\hline
$h/(x+1)$ & $f_1f_1^*f_2f_2^*$   & $f_1f_1^*f_3f_3^*$ &  $f_2f_2^*f_3f_3^*$\\ \hline
$f$ & $f_3$ & $f_2$& $f_1$ \\ \hline
$d_R$ &3&9&27 \\ \hline
$d_T$ &2&2&2 \\ \hline
$\min(2d_T,d_R)$&3&4&4 \\ \hline
\end{tabular}
\label{table6}

\caption{ Length $27$ continued}
\end{table}
\item[$n=29$] Two cyclic self dual codes with $h=X+1$ and $$f,\, g=X^{14} + \w X^{13} + \w X^{11} +\w^2 X^{10} + X^9 + \w^2 X^8 + \w x^7 + \w^2X^6 + X^5 +
        \w^2 X^4 +\w X^3 + \w X + 1.$$

\item[$n=31$]$$X^{31}+1=(X+1) \prod_{j=1}^3 f_j f_j^*,$$
with $$f_1=X^5+X^2+1,\,f_2=X^5+X^3+X^2+X+1,\,f_3=X^5+X^2+X+1$$%
\begin{table}[ht]
\small
\begin{tabular}{|l|l|l|l|l|l|l|l|l|l|l|}
\hline
$h/(x+1)$ &$1$ & $1$ & $1$ &$1$ & $f_1f_1^*$ &$f_1f_1^*$ &$f_2f_2^*$  &$f_2f_2^*$  &$f_3f_3^*$& $f_3f_3^*$ \\ \hline
$f$ &$ f_1f_2f_3$ & $f_1f_2f_3^*$&$ f_1f_2^*f_3$& $ f_1f_2^*f_3^*$& $f_2f_3$  & $f_2f_3^*$&$f_1f_3$  & $f_1f_3^*$&$f_1f_2$  & $f_1f_2^*$\\ \hline
$d_R$ &6&6&6& 6&8&8&6 &6&6&6\\ \hline
$d_T$&6&6&6& 6 &4&4& 4&4&5&5\\ \hline
$\min(2d_T,d_R)$&6&6&6&6  &8&8&6 &6&6&6\\ \hline
\end{tabular}
\label{table7}

\caption{ Length $31$}
\end{table}
\begin{table}[ht]
\small
\begin{tabular}{|l|l|l|l|}
\hline
$h/(x+1)$ & $f_1f_1^*f_2f_2^*$   & $f_1f_1^*f_3f_3^*$ &  $f_2f_2^*f_3f_3^*$\\ \hline
$f$ & $f_3$ & $f_2$& $f_1$ \\ \hline
$d_R$ &10&8&6 \\ \hline
$d_T$ &2&3&3 \\ \hline
$\min(2d_T,d_R)$&4&6&6 \\ \hline
\end{tabular}
\label{table8}

\caption{ Length $31$ continued}
\end{table}
\end{itemize}
\section{Conclusion}

In this article, we have derived the theory of cyclic codes over the noncommutative ring of matrices of order $2$ over $\F_2.$
In particular we have given a characterization of cyclic codes and their duals as right ideals in terms of two generators.
We have proved the existence of infinitely many nontrivial cyclic codes for the Euclidean product. Their Bachoc images are
formally self dual quaternary codes. All this was derived of the case of odd length codes. A natural question is the generalization
to even length. In view of the known results \cite{DL} for the ring $\Z_4$ this might be difficult. A worthwhile motivation for this effort would be to construct hermitian self dual codes that could lead to lattices by the construction in \cite{B}.

\end{document}